\documentclass{article}
\usepackage[utf8]{inputenc}
\DeclareUnicodeCharacter{0300}{\`{}} % Example for combining grave accent

\usepackage[T1]{fontenc}
\usepackage{authblk}
\usepackage{setspace}
\usepackage[margin=1.25in]{geometry}
\usepackage{float}
\usepackage{graphicx}
\usepackage{subcaption}
\usepackage{amsmath}
\usepackage{lineno}

\usepackage[style=nejm, 
citestyle=numeric-comp,
sorting=none]{biblatex}
\title{Leveraging Knowledge Distillation for Lightweight Skin Cancer Classification: Balancing Accuracy and Computational Efficiency}

\author[1]{Niful Islam}
\author[2,3]{Khan Md Hasib}
\author[4]{Fahmida Akter Joti}
\author[5*]{Asif Karim}
\author[5]{Sami Azam}

\affil[1]{Department of Computer Science and Engineering, United International University, United City, Madani Avenue, Badda, Dhaka 1212, Bangladesh.}
\affil[2]{Department of Computer Science and Software Engineering, The University of Western Australia, Perth, WA 6009, Australia.}
\affil[3]{Department of Computer Science and Engineering, Bangladesh University of Business and Technology, Dhaka 1216, Bangladesh.}
\affil[4]{Ahsanullah University of Science and Technology, Dhaka 1208, Bangladesh.}
\affil[5]{Charles Darwin University, NT 0810, Australia.}

\affil[*]{Address correspondence to: asif.karim@cdu.edu.au}

\date{} % Suppress the date
\begin{document}

\maketitle

\begin{abstract}
Skin cancer is a major concern to public health, accounting for one-third of the reported cancers. If not detected early, the cancer has the potential for severe consequences. Recognizing the critical need for effective skin cancer classification, we address the limitations of existing models, which are often too large to deploy in areas with limited computational resources. In response, we present a knowledge distillation based approach for creating a lightweight yet high-performing classifier. The proposed solution involves fusing three models, namely ResNet152V2, ConvNeXtBase, and ViT Base, to create an effective teacher model. The teacher model is then employed to guide a lightweight student model of size 2.03 MB. This student model is further compressed to 469.77 KB using 16-bit quantization, enabling smooth incorporation into edge devices. With six-stage image preprocessing, data augmentation, and a rigorous ablation study, the model achieves an impressive accuracy of 98.75\% on the HAM10000 dataset and 98.94\% on the Kaggle dataset in classifying benign and malignant skin cancers. With its high accuracy and compact size, our model appears to be a potential choice for accurate skin cancer classification, particularly in resource-constrained settings.
\end{abstract}

%%%%%% Main Text %%%%%%
\section{Introduction}

The skin, the largest organ of the human body, protects the internal organs from external hazards such as dust, heat, polluted water, and ultraviolet rays. However, the primary line of defense is frequently exposed to external agents causing several skin diseases including cancer \cite{roger2022neuroimmune}. Over the past few years, the number of skin cancer patients has grown significantly, accounting for one-third of the reported cancers \cite{zia2022classification, aladhadh2022effective}. Benign and malignant skin cancers are two distinct types with contrasting characteristics \cite{naqvi2022benign}. Benign skin cancers typically have a modest growth rate and remain confined, without spreading to other tissues. They are usually not life-threatening, though they could be uncomfortable if left untreated \cite{gouda2022detection}. Malignant skin cancers, on the other hand, are more aggressive in their nature. They could infiltrate adjacent tissues and, if they are not found and treated promptly, spread to other areas of the body, so becoming a major health hazard \cite{bratchenko2022classification}. Skin tumors that are malignant frequently have uneven borders and proliferate quickly. Leveraging computer vision-based technologies for the early detection of skin cancer offers the potential for enhanced efficiency and accuracy compared to traditional approaches.

Traditional image classification techniques primarily rely on Convolutional Neural Networks (CNNs). Convolutional layers and the pooling layers are the essential elements of CNNs. While convolutional layers effectively extract useful features from the input image, the pooling layer downsamples the spatial dimensions of the feature map, conserving the important features and enhancing computational efficiency \cite{pi2022broadband}. Although the CNNs excel in a variety of image recognition tasks, the networks require high computational power and a large number of labeled training data which is often expensive to obtain. Transfer learning offers a solution to this challenge by enabling the utilization of pretrained neural networks trained on extensive datasets containing generic images. Throughout the training process, the model learns to extract useful features that can be leveraged in the related tasks. Since a new task begins with a pretrained CNN backbone, it requires only the top layers to be fine-tuned which significantly reduces the computational burden \cite{iman2023review}. Furthermore, transfer learning also allows the model to work with less training data since only a small section of the model needs to be trained.  

Another restriction of CNNs is their limited ability to accommodate global context. Since CNNs incorporate convolutional layers that are limited to their kernel size, the networks fail to capture larger and more global patterns beyond the receptive field of the kernel. Vision Transformer (ViT) \cite{dosovitskiy2020image} has emerged as a solution to this problem. The ViT, originated from the transformer \cite{vaswani2017attention} architecture, incorporates self-attention mechanism. Dissimilar to CNNs, ViT's self-attention mechanism captures long-range dependencies and global context from the input image. The transformer devices an input image into non-overlapping segments named patches which are treated as sequence tokens and processed using self-attention layers. By doing so, ViT overcomes CNNs' shortcomings in handling global context, making it especially useful for applications that require comprehending relationships across all segments are essential. The use of Vision Transformer has yielded encouraging results in a variety of computer vision applications, outperforming state-of-the-art CNN architectures \cite{shamshad2023transformers, amjoud2023object, thisanke2023semantic}. 

The growing size and complexity of deep learning models present a substantial challenge, especially when deployed on resource-constrained devices. As these models grow in size, their computational and memory needs increase, making them unsuitable for practical usage on devices with limited capabilities, such as mobile phones or edge devices \cite{islam2023toward, beyer2022knowledge}. To solve this issue, a mechanism has arisen for compressing these bulky models into more efficient alternatives while maintaining performance. This procedure entails training a smaller model, commonly known as the student model, to match the predictions of a bigger, pre-existing model, known as the teacher model. The student model learns from both labeled data and the knowledge embedded in the softer probability distributions provided by the teacher model \cite{song2022spot}. This information transmission not only allows for the design of more compact models but also improves their generalizability \cite{xu2023teacher}. The ability to deploy these streamlined models on resource-constrained devices makes deep learning applications more accessible and useful in real-world scenarios.

In response to the demand for precise skin cancer classification using a lightweight classifier suitable for integration into edge devices, we have introduced a knowledge-distillation approach. The presented system incorporates a six-stage image preprocessing technique along with data augmentation to enhance both the quality and quantity of images. Furthermore, we have constructed a robust teacher model, consisting of a vision transformer and two CNN classifiers, to guide the small student model. Subsequently, the student model, with a size of 2.03 MB, is trained, utilizing only three convolutional layers for classification. We have conducted an extensive ablation study on the student model for finding the best hyperparameters. The model is further compressed using 16-bit quantization, resulting in the final size of the classifier being 469.77 KB. We have compared the performance with existing results, and the proposed model demonstrates superiority over prior works. To summarize, this research has the following major contributions.
\begin{itemize}
    \item We have introduced a six-stage image preprocessing technique, designed to highlight the lesion regions in skin cancer images.
    \item We have developed a fusion model comprising Vision Transformer, ResNet152V2, and ConvNeXtBase, which incorporates channel attention with CNN feature extractors to enhance performance. This robust teacher model is utilized to train the student model.
    \item This research article offers a lightweight student model with only three convolutional layers. The model has only 0.16 million parameters, occupying a compact size of 2.03 MB. Additional compression through quantization reduces the model to 469.77 KB in size.
    \item The proposed model achieves an accuracy of 98.94\% and 98.75\% on the Kaggle and HAM1000 datasets, respectively, for accurately classifying benign and malignant skin cancer images. A comparison with existing research demonstrates the superiority of the proposed method. 
\end{itemize}
The remaining article is structured as follows. In Section \ref{sec:rw}, we delve into the current approaches for classifying skin cancer images, examining their limitations, and identifying a research gap that this article aims to fill. Subsequently in Section \ref{sec:method}, we have thoroughly explained the proposed method followed by the outcomes of the method in Section \ref{sec:results}. Lastly, the article terminates in Section \ref{sec:conclusion}.

\section{Related Works}
\label{sec:rw}
Lately, considerable attention has been directed towards achieving high-accuracy classification of skin cancer. Among them, the CNN-based methods have emerged as the most frequently employed techniques. Ali et al. \cite{ali2022multiclass} investigated eight EffecientNets (B0-B7) for effectively classifying skin cancer. The proposed approach accommodated an image preprocessing pipeline that consisted of three sequential stages: hair removal, data augmentation, and image resizing. Following the image preprocessing step, they fine-tuned the EffecientNets with different hyper-parameters. While fine-tuning EffecieneNet B0-B5 only required training the top layers, made of fully connected neural networks, the training process of EffecientNet B6-B7 incorporated two steps. In the first step, the top layer was trained while keeping the convolutional feature extractors frozen. In the second step, the last four convolutional blocks were trained, keeping the rest of the model frozen. Moreover, EffecientNet B0-B5 were trained with Stochastic Gradient Descent (SGD) optimizer while EffecientNet B6 and B7 were trained on Adam optimizer since Adam produced better performance on the larger models. Among the eight EfficientNet models, EffecientNet B4 produced the best performance with an accuracy of over 87\%. This research work, however, does not present any ablation study leaving the impact of the hyperparameters unknown. Mridha et al. \cite{mridha2023interpretable} presented a lightweight model having over 5 million parameters. The model adeptly extracted features by employing only two convolutional blocks, each consisting of two convolutional layers along with one max pooling layer and a dropout layer. The system incorporated Grad-CAM and Grad-CAM++ for visualizing the attention heatmaps. Nevertheless, this model failed to achieve a higher accuracy due to its lightweight nature. Zia et al. \cite{zia2022classification} experimented with MobileNetV2 and DenseNet201, two lightweight image classifiers. In that research, the authors presented two modified classifiers with MobileNetV2 and DenseNet201 backbone which followed three additional convolutional layers for extracting additional features. According to the experiment, the fine-tuned models significantly outperformed the base models. Moreover, the performance of DenseNet201 exceeded MobileNetV2. The authors also employed Grad-CAM for visualizing the attention heatmap of the classifier. This research, however, only leveraged one dataset. Gururaj et al. \cite{gururaj2023deepskin} employed various image preprocessing techniques along with two state-of-the-art image classifiers for classifying skin cancer images from HAM-1000 dataset. The image preprocessing techniques included both oversampling and undersampling to address the data imbalance problem, Dull Razor method for removing noise, and segmentation of the lesion regions with an autoencoder. Finally, the images were classified using DenseNet169 and ResNet50. The best performance was achieved by DenseNet169 with undersampling. Since the system required an image to pass through two CNN models (i.e., autoencoder and DenseNet169), the computational complexity is very high.

Recently transformer-based models have gained attention due to their high classification performance. Xin et al. \cite{xin2022improved} modified Vision Transformer (ViT) by integrating multiscale an overlapping sliding window. Furthermore, the improved transformer model incorporated multi-scale patch embedding. The proposed ViT was evaluated on HAM1000 and a custom dataset where the model achieved over 94\% accuracy. Despite the sliding window mechanism, the self-attention process consumed very high computational resources making the model slow for high-resolution images. Ayas \cite{ayas2023multiclass} leveraged Swin Transformer for effectively classifying skin cancer images. Swin Transformer \cite{liu2021swin} is a modified Vision Transformer capable of producing high classification accuracy with the help of shifted window mechanism along with hierarchical feature extraction. According to the research, the Swin Transformer was able to achieve an outstanding performance in differentiating eight types of skin cancers. While there were smaller Swin Transformer models available, the best performance was produced by Large\_22K which was responsible for high resource consumption. Since the feature extraction process of CNN and transformer is different, combining these two results in more non-overlapping features which leads to higher performance. Therefore, Hao et al. \cite{hao2023convnext} proposed a fusion model named ConvNeXt-ST-AFF that fuses a CNN model (ConvNeXt) and a transformer model (Swin Transformer). This architecture effectively extracts useful features from two state-of-the-art models and combines them for final prediction. Similar to the previous models, this achieves a high performance by consuming high resources.

Since single model classifiers sometimes fail to achieve significant accuracy, ensembling multiple classifiers produces better performance. Imran et al. \cite{imran2022skin} fused three deep learning models namely CapsNet, VGG and ResNet for constructing an enable model. Although the model achieved an outstanding performance of 93.5\% accuracy, surpassing recent works, its heavyweight nature made it an impractical option for low-constraint devices. Keerthana et al. \cite{keerthana2023hybrid} introduced a fusion model composed of two pre-trained deep neural networks. The features extracted from these networks were concatenated and subsequently fed into a machine learning model for the final classification. Hasan et al. \cite{hasan2022dermoexpert} presented a stacked ensemble method named dermoexpert. This model extracted features using three CNN models. In the subsequent stage, the features from the first two CNN streams were fused to form a fourth stream. The fusion process included fusion by channel-averaging and fusion by channel concatenation. Following the second stage, a series of fully connected blocks were integrated with the streams. Finally, the prediction of the four streams was combined using soft voting for the final classification. Similar to the previous works, this research also consumed very high computational resources.

\begin{table}[H]
    \centering
    \caption{Major contributions and limitations of recent studies on skin cancer}
    \begin{tabular}{p{3cm} p{2cm} p{3.9 cm} p{3.9cm}}
    \hline
        \textbf{Paper} & \textbf{Dataset} & \textbf{Contribution} & \textbf{Limitation} \\ \hline
        Ali et al. \cite{ali2022multiclass} & HAM10000 & A comparison is made among eight EffecientNets. & No ablation study presented. \\ \\ 
        Mridha et al. \cite{mridha2023interpretable} & HAM10000 & Presented an interpretable model. & Relatively lower classification accuracy. \\ \\
        Zia et al. \cite{zia2022classification} & Kaggle & Presents a novel fine-tuning approach of MobileNet and DenseNet. & Leverages only one dataset. \\ \\
        Gururaj et al. \cite{gururaj2023deepskin} & HAM10000 & Very high classification performance. & Computational complexity is also high. \\ \\
        Xin et al. \cite{xin2022improved} & HAM10000, Private dataset & A novel transformer architecture with very high performance. & The model is very slow for high-resolution images. \\ \\
        Ayas \cite{ayas2023multiclass} & ISIC 2019 & Achieves very high performance. & Consumes high computation. \\ \\
        Hao et al. \cite{hao2023convnext} & ISIC 2019, Private dataset & Presents a novel architecture made of state-of-the-art models. & Consumes high computation. \\ \\
        Imran et al. \cite{imran2022skin} & ISIC & A fusion model with very high performance & The heavyweight nature makes it difficult to use for low-constraint devices. \\ \\
        Keerthana et al. \cite{keerthana2023hybrid} & ISBI 2016 & Presents a detailed comparison of state-of-the-art image classifiers & Consumes high computational resources. \\ \\
        Hasan et al. \cite{hasan2022dermoexpert} & ISIC 2016, ISIC 2017, ISIC 2018 & Presented a stacked enseble model that achieved very high classification performance. & Very high computaional resource is required. \\ \hline
    \end{tabular}
    \label{tab:rw-comp}
\end{table}

A summary of the recent works discussed in this section is presented in Table \ref{tab:rw-comp}. According to the analysis, the majority of the research works achieved a high classification performance by compromising computational cost. Hence, this research strikes a balance between classification accuracy and resource efficiency. 

\section{Proposed Method}
\label{sec:method}
\begin{figure}[h]
    \centering
    \includegraphics[height=10cm, width=1.05\linewidth]{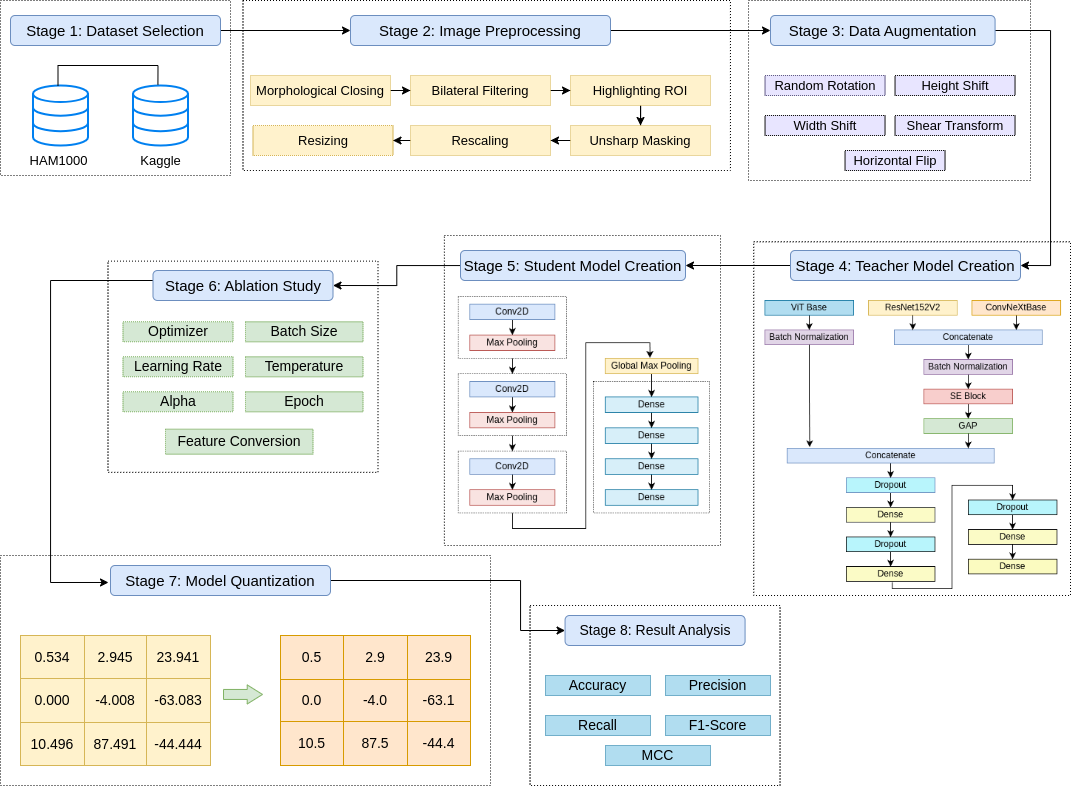}
    \caption{Overview of the proposed method.}
    \label{fig:flow-diagram}
\end{figure}

As presented in Figure \ref{fig:flow-diagram}, the proposed solution consists of eight stages. This section illustrates the first seven stages in detail. 

\subsection{Dataset Details}
In the evaluation of the proposed model, we have utilized two benchmark datasets. The first is the Human Against Machine 10000 dataset, commonly referred to as HAM10000 \cite{tschandl2018ham10000}, and the second is the Kaggle dataset \cite{kaggle_skin_cancer}. The HAM10000 consists of 10,015 images distributed across seven classes. We have categorized these seven classes into two groups: benign and malignant for classification. In the dataset, classes such as Melanoma, Basal cell carcinoma, Actinic keratosis, Benign keratosis-like lesions, and Vascular lesions are categorized as malignant. On the other hand, Melanocytic nevi and Dermatofibroma are classified as benign. Notably, the Kaggle dataset, encompassing 3,297 images, already includes pre-categorized images as benign or malignant. While the images in the HAM10000 dataset are of size 450$\times$600, the images in the Kaggle dataset are already resized to 224$\times$224. Table \ref{tab:dataset} presents a class-wise distribution of the number of images. For model training and evaluation, we have divided the dataset into a ratio of 70:10:20 for training, validation, and testing, respectively.

\begin{table}[h]
    \centering
    \caption{Class-wise distribution of the images.}
    \begin{tabular}{p{3cm} p{3cm} p{3cm}}
    \hline
        \textbf{Category} & \textbf{HAM10000} & \textbf{Kaggle} \\ \hline
        Benign & 6,820 & 1,800 \\ 
        Malignant & 3,195 & 1,497 \\ 
        Total & 10,015 & 3,297 \\\hline
    \end{tabular}
    \label{tab:dataset}
\end{table}

\subsection{Image Preprocessing}
\begin{figure}[h]
    \centering
    \includegraphics[height=10cm, width=1.05\linewidth]{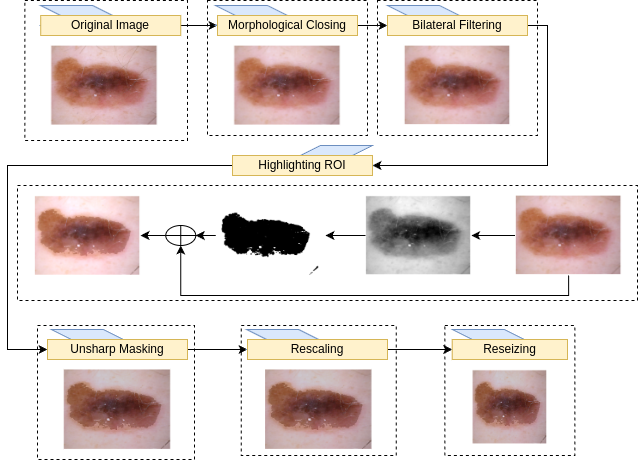}
    \caption{Image preprocessing steps.}
    \label{fig:preprocess}
\end{figure}

The image preprocessing stage comprises six steps aimed at enhancing image quality. These techniques involve noise removal, highlighting the lesion region, and resizing images to accommodate the pretrained image classifiers. Figure \ref{fig:preprocess} depicts the image preprocessing stages in detail. 

\subsubsection{Morphological Closing}
Since skin cancer images contain unnecessary hairs which often hinder the classification process, morphological closing is applied at the very first stage of the image preprocessing. Morphological closing is a mathematical operation that involves dilation followed by erosion. It removes small gaps while preserving the original structure in an image \cite{lu2021data}. The morphological closing operation can be denoted by equation \ref{eq:morphological-closing}. In this equation, $I$ denotes the input image and $B$ is the structuring element. A structuring element is a kernel that guides neighborhood operations like dilation and erosion. For the proposed method, a structuring element of size 5$\times$5 is selected to strike a balance between the efficiency of hair removal and the risk of leaving behind small remnants. The higher kernel size ensures the removal of moderately sized hair strands. In this operation, the input image is first dilated with the structural element. The dilation process, denoted by $\oplus$, causes objects in the image to expand in size, filling in small gaps. Subsequently, erosion, denoted by $\ominus$, is applied to the resulting image for shrinking the boundaries of regions. 

\begin{equation}
    \label{eq:morphological-closing}
    I \cdot B = (I \oplus B) \ominus B
\end{equation}

\subsubsection{Bilateral Filtering}
Skin cancer images typically contain noise that is challenging to eliminate with conventional low-pass filters, since they may remove subtle diagnostic features. Therefore, we have employed bilateral filter to remove noise while preserving fine details. Unlike standard smoothing filters, which blur the entire image uniformly, bilateral filtering takes into account both the spatial distance between pixels and the intensity difference \cite{li2021multimodal, khattar2022computer}. It operates by averaging the intensity of surrounding pixels and assigning additional weights to pixels that are spatially close and have similar intensities. The mathematical equation of bilateral filtering is expressed in Equation \ref{eq:bf}. For an input image $I$, the filtered intensity value at position $q$ can be obtained by considering both the spatial distance, $\|p - q\|$, and the intensity differences, $|I_p - I_q|$ from the central pixel $p$ and its neighborhood pixels $q$ in the spatial neighborhood $S$. While measuring the distance, the spatial Gaussian filter, $G_{\sigma_s}$, and the range Gaussian filter, $G_{\sigma_r}$ ensures that neighboring pixels with comparable intensity have a higher influence on the filtered image. Finally, the output is normalized with $W_p$ to prevent over-amplification.

\begin{equation}
\label{eq:bf}
BF[I]_p = \frac{1}{W_p} \sum_{q \in S} G_{\sigma_s}(\|p - q\|) \cdot G_{\sigma_r}(|I_p - I_q|) \cdot I_q
\end{equation}

\subsubsection{Highlighting ROI}
Segmenting the lesion regions from the skin cancer images and leveraging the segmented portion for classification typically results in a better performance \cite{bibi2022skin, monika2020skin}. However, incorporating a segmentation model like U-Net remarkably increases the computational burden. On the other hand, relying on traditional segmentation techniques might produce inconsistent results. Therefore, the lesion regions in the skin cancer images are highlighted to ensure feature preservation and prioritize relevant information. This step encompasses three key phases: grayscale conversion, image segmentation, and linear blending. 

We have leveraged Otsu's thresholding technique to segment the image which typically works on grayscale images. Hence, the input RGB image is converted into a grayscale image using a linear equation described in Equation \ref{eq:gray}. In this equation, the red, green, and blue channels are described with $R$, $G$, and $B$ respectively. The gray image, $I_g$, is computed by multiplying each pixel value with a predefined constant.

\begin{equation}
\label{eq:gray}
    I_g = 0.299 \cdot R + 0.587 \cdot G + 0.114 \cdot B
\end{equation}

Following the grayscale conversion, the image is then segmented with Otsu's method. Otsu's method calculates a threshold value based on the pixel intensity of the image which is leveraged for binary segmentation \cite{peng2023multi}. For measuring the optimal threshold value, it iterates over the potential threshold values, and the value having the maximizing inter-class variance is returned as the threshold. The formula for inter-class variance is calculated using Equation \ref{eq:otsu}. In this equation, $\omega_{0}(t)$ and $\omega_{1}(t)$ represent the probabilities of two classes separated by a threshold $t$. The equations of $\omega_{0}(t)$ and $\omega_{1}(t)$ are presented in Equation \ref{eq:w0} and \ref{eq:w1} respectively. $\omega_{0}(t)$ is the cumulative probabilities of pixel intensities up to the threshold $t$ while $\omega_{1}(t)$ measures the probabilities from $t+1$ to $L-1$, where $L-1$ is the highest intensity level. In Equation \ref{eq:otsu}, $\sigma_{0}$ and $\sigma_{1}$ are the variances of two classes.

\begin{equation}
\label{eq:otsu}
    \sigma_{w}^{2}(t) = \omega_{0}(t) \sigma_{0}^{2}(t) + \omega_{1}(t) \sigma_{1}^{2}(t)
\end{equation}

\begin{equation}
\label{eq:w0}
    \omega_{0}(t) = \sum_{i=0}^{t} P(i)
\end{equation}

\begin{equation}
\label{eq:w1}
    \omega_{1}(t) = \sum_{i=t+1}^{L-1} P(i)
\end{equation}

Following the selection of the threshold value using Otsu's method, the image is segmented using a binary thresholding mechanism in which pixel intensities less than the threshold are assigned to one class, and those greater than the threshold are assigned to another class. Equation \ref{eq:segment} illustrates the segmentation process in detail. In this equation, $I$ denotes the input image and $I_s$ denotes the segmented image and $x$, $y$ are two random positions in the image. 

\begin{equation}
\label{eq:segment}
I_s(x, y) =
\begin{cases} 
    1 & \text{if } I(x, y) \geq t \\
    0 & \text{if } I(x, y) < t
\end{cases}
\end{equation}
Now the segmented image is linearly blended with the input image, resulting in a highlighted lesion region. The linear blending mechanism is illustrated in Equation \ref{eq:linear_blending}, where $I_s$ is the segmented image and $\alpha$ is a constant which is 1.14 for this experiment and $I$ is the input image. The resulting highlighted image is denoted with $I_h$. 

\begin{equation}
\label{eq:linear_blending}
    I_h = I + \alpha \times I_s
\end{equation}

\subsubsection{Unsharp Masking}
Unsharp masking is an image sharpening method that employs low pass filters to sharpen details and improve overall visibility \cite{song2022unsharp}. It improves skin cancer images by accentuating high-frequency components corresponding to edges and tiny details. The image sharpening algorithm has three major steps. In the first step, the input image is convoluted with a low pass filter. Let $I$ define the input image and $f$ be a low pass filter. Now the resulting blurred image, $I_b$ is obtained by operation illustrated in Equation \ref{eq:bluring}. The selection of the low pass filter plays a crucial role in this step. For the proposed solution, the kernel is automatically determined from the standard deviation values of 2.0 in both the X and Y directions. 

\begin{equation}
     \label{eq:bluring}
         I_b = I * f
\end{equation}

The second step involves extracting the high frequency components by subtracting the blurred image from the input image. Equation \ref{eq:masking} exhibits the mathematical process. In this process, $I$ resembles the original image and $I_b$ is the blurred image obtained from the previous step. The resulting mask, $I_m$, encapsulates high frequency elements and the sharp edges in the input image.

\begin{equation}
     \label{eq:masking}
         I_m = I - I_b
\end{equation}

Lastly, as described in Equation \ref{eq:adding_mask}, the mask is amplified by multiplying it with a constant, $k$, and blended with the input image. This process is identical to linear blending. For the proposed solution, the value of $k$ is selected as 1.2.

\begin{equation}
\label{eq:adding_mask}
    I_{um} = I + k \times I_m
\end{equation}

\subsubsection{Rescaling}
Rescaling is a commonly used preprocessing technique in computer vision. This technique converts pixel intensity values originally ranging from 0 to 255 into a normalized scale of 0 to 1. Equation \ref{eq:rescale} illustrates the rescaling process where $I$ denotes the original image and $I_r$ is the rescaled image. While this stage may not introduce any visual changes to the image, rescaling facilitates faster convergence, leading to reduced computation.
\begin{equation}
\label{eq:rescale}
I_r = I/255
\end{equation}

\subsubsection{Resizing}
Image resizing is also another frequently employed image preprocessing technique, particularly in the context of transfer learning. Typically state-of-the-art image classifiers are trained on ImageNet, that have a resolution of 469$\times$387 \cite{deng2009imagenet}. However, they are typically resized to 255$\times$255 or 224$\times$244 \cite{magdy2023performance}. Therefore, resizing the images to 224$\times$224 typically yields a better performance. Moreover, lowering the image size significantly reduces the computation. Since the images in the Kaggle dataset are already resized to the specified size, only the images from the HAM1000 dataset undergo this stage.

\subsubsection{Effects of Image Preprocessing}
\begin{figure}
    \centering
    \includegraphics[height=6cm, width=1\linewidth]{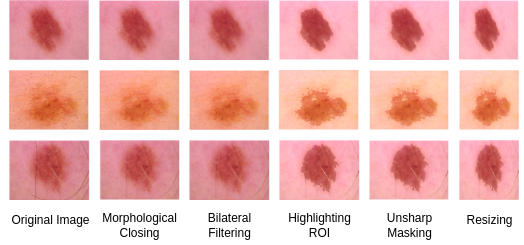}
    \caption{Changes in the input image after each preprocessing step.}
    \label{fig:before-after-preprocessing}
\end{figure}

The changes in the input image after each preprocessing step are presented in Figure \ref{fig:before-after-preprocessing}. Since rescaling does not bring any visual changes to the images, the impact of this step is excluded. The figure shows that the hairs present in the image are removed in the first step. Subsequently, the noise is eliminated. In the third step, the region of interest, specifically the lesion region, becomes more distinctly visible. Following that, unsharp masking highlights the edges in the lesion region, making the decision-making process easier. Finally, resizing ensures the image is compatible with the pretrained image classifiers.

\subsection{Data Augmentation}
Various data augmentation techniques have been employed to mitigate overfitting, applied to both the teacher and student models. The augmentations have been applied to both the teacher model and the student models. The augmentation techniques that have been applied to the training dataset are random rotation, height shift, width shift, shear transform, and horizontal flip. Random rotation creates variations by spinning images within a specific range; height and width shifts change the position of items; shear transform skews the image; and horizontal flip horizontally reflects images. These augmentations imitate real-world variability, allowing the model to learn invariant traits and perform better on previously unseen data.

\subsection{Teacher Model Creation}
\begin{figure}[h]
    \centering
    \includegraphics[height=8.5cm, width=1\linewidth]{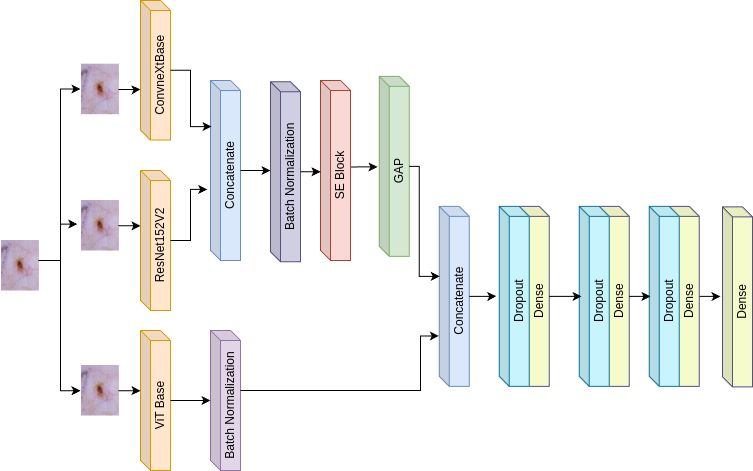}
    \caption{Teacher model used for knowledge distillation.}
    \label{fig:teacher}
\end{figure}

We have constructed a robust teacher model made of three feature extractors for guiding the lightweight student model. As presented in Figure \ref{fig:teacher}, the proposed teacher model is composed of two streams. The first stream is composed of two CNN feature extractors: ResNet152V2 \cite{he2016deep} and ConvNeXtBase \cite{liu2022convnet}, two state-of-the-art image classifiers. Since the feature extraction process of transformer architecture is different from CNNs, the second stream incorporates ViT Base \cite{dosovitskiy2020image}, a robust transformer network. The features from these streams are fused to construct a strong teacher model. 

ResNet152V2, a variant of the ResNet architecture, is specifically engineered to mitigate the challenge of the vanishing gradient problem. This issue arises during the training of deep networks, where gradients diminish significantly during backpropagation, leading to slowed or stalled learning progress. Residual networks address this issue through the integration of residual blocks. A residual block creates a shortcut or residual connection by adding the block's input to its output. The architectural enhancement allows the network to learn the residual mapping, which facilitates the model's ability to detect gradients and identity mappings during training. Due to the fine gradient preservation, it is possible to create deep neural networks without encountering the vanishing gradient problem. ResNet152V2 extends the original ResNet architecture by incorporating batch normalization that speeds up the training process \cite{hwooi2022deep}. This model consists of 152 layers that effectively extract useful features from the image, making the model a reliable classifier.

ConvNextBase is a unique vision model that modernizes the ResNet block toward high efficiency and accuracy. The core innovation of the transformer-inspired architecture is its ConvNeXt block. Instead of relying on the traditional residual blocks, the model employs inverted bottleneck block, commonly seen in transformer architecture. The inverted bottleneck block integrates a depth-wise separable convolution followed by a pointwise convolution. Moreover, a layer normalization along with the Gaussian Error Linear Unit (GELU) activation function further enhances the model's classification ability. With other micro design changes, the model is proven to be one of the most powerful classifiers in the era of transformers. 

Derived from the Transformer network, the Vision Transformer (ViT) has become a formidable asset in the realm of computer vision. Operating by segmenting input images into non-overlapping patches, ViT then projects these patches linearly through multiplication with a trainable weight matrix. Following this, position embedding is incorporated with the patches and forwarded to the transformer block, encompassing layer normalization, multi-head self-attention, and multi-layer perception (MLP). After a series of transformer blocks, the images are classified with an MLP head.

The first stream of the proposed teacher model concatenates the features extracted from the pretrained ResNet152V2 and ConvNeXtBase. The combined features are then passed to batch normalization and Squeeze and Excitation (SE) block. SE block is an attention mechanism that prioritizes the channels that are more important for classification \cite{hu2018squeeze}. The block consists of three operations: squeezing, exciting, and recalibrating. In the squeeze operation, illustrated in Equation \ref{eq:squeeze}, the spatial dimension of a channel is reduced with global average pooling. Let $X_c$ denote a feature map with $H$ height and $W$ width. Now, the summary of the channel, denoted with scaler $Z_c$, is obtained by taking the average of the channel $X_c$.

\begin{equation}
\label{eq:squeeze}
    z_c= \frac{1}{H \times W}\sum_{i=1}^{H}\sum_{j=1}^{W}X_c(i,j)
\end{equation}

In the excitation phase, presented in Equation \ref{eq:excitation}, the scaler is passed through two fully connected layers. Let $W_1$ denote the weights of the first layer and $W_2$ be the weights of the second layer. The first layer is active with ReLU and the second layer is activated with sigma. The excitation phase returns the channel weight $s_c$ which is multiplied with the feature map to get the recalibrated feature map $y_i$. Equation \ref{eq:reclaibration} illustrates the recalibration process.
\begin{equation}
\label{eq:excitation}
    s_c=\sigma(W_2\times ReLU(W_1\times z))
\end{equation}

\begin{equation}
\label{eq:reclaibration}
    y_i= s_c \times X_c
\end{equation}

Since the pretrained classifiers are trained on ImageNet, which is significantly different from biomedical images, an attention mechanism greatly enhances the performance. Following the SE block, a Global Average Pooling (GAP) layer is incorporated to transform the 2D feature maps into a 1D feature vector. 

The second stream, on the other hand, consists of a ViT Base and a batch normalization layer. The features extracted from the first stream and the second stream are combined to construct the final fusion model. The fusion model then follows three blocks of dropout and dense layer. The first two dropout layer randomly drops 50\% neurons and the last layer drops 20\% neurons. The dropout layers prevent overfitting. The first three dense layers, on the other hand, consist of 256, 128, and 64 neurons which are activated with Scaled Exponential Linear Unit (SELU). The SELU activation function, known for its simplicity and effectiveness in preventing overfitting, is utilized in the final layer, comprising two neurons employing the softmax activation function \cite{verma2023revisiting}. The final layer comprises two neurons utilizing the softmax activation function. The model has over 236 million parameters out of which 3.39 million parameters are trainable. The size of the model is 975.85 megabytes.

\subsection{Student Model Creation}
\begin{figure}[h]
    \centering
    \includegraphics[height=4.3cm, width=1\linewidth]{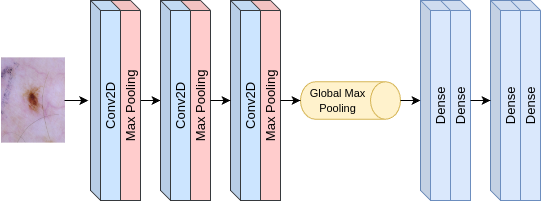}
    \caption{Student model used for classification.}
    \label{fig:student}
\end{figure}

The student mode, presented in Figure \ref{fig:student}, is composed of three convolutional blocks. Each block consists of a 3$\times$3 convolutional layer and a 2$\times$2 max pooling layer. The initial convolutional layer has 32 filters. Similarly, the second and the third convolutional layer employs 64 and 128 layers respectively. Increasing the number of filters in the proposed network improves its ability to capture complex hierarchical patterns in the input image, resulting in a more nuanced representation. The max pooling layer effectively reduces the spatial dimension of the image, retaining the most prominent features. The output of the final pooling layer is then subjected to global max pooling to generate a condensed representation of the spatial information. Subsequently, the feature vectors are passed through three fully connected layers with SELU activation having 256, 128, and 32 neurons. The final layer consists of 2 neurons with softmax activation for classification. Since the teacher model is regularized with dropout layers, no dropout layers are integrated into this model. 

\begin{figure}[h]
    \centering
    \includegraphics[height=7.3cm, width=0.9\linewidth]{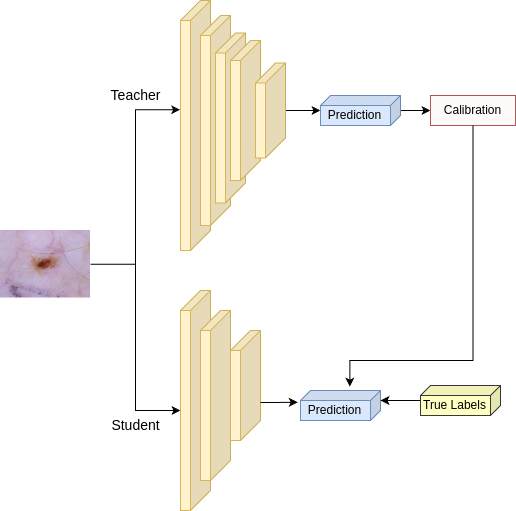}
    \caption{Knowledge distillation process.}
    \label{fig:knowledge-distillation}
\end{figure}

The model is trained with knowledge distillation, a technique in which a simple model, known as the student, is trained to replicate the predictions of a more complex model, known as the teacher model \cite{islam2023toward}. Figure \ref{fig:knowledge-distillation} presents an overview of the knowledge distillation process. The approach involves sharing knowledge from the teacher to the student through softened probabilities, which are adapted using two key hyperparameters: temperature and alpha. The temperature parameter regulates the level of softening, which influences the smoothness of the probability distribution, whilst the alpha parameter sets the weighting of hard and soft targets throughout the training phase. This enables the student model to benefit from the rich knowledge stored in the teacher's predictions, resulting in a more efficient and accurate learning process.

\subsection{Ablation Study}
In order to determine the optimal hyperparameters, an extensive ablation study has been conducted on the student model. The outcomes of the study on the HAM10000 and the Kaggle datasets are presented in Table \ref{tab:ham-ablation} and \ref{tab:kaggle-ablation} respectively. The results depict a similar impact of the seven hyperparameters. According to the analysis, the feature conversion mechanism is the most influential one among the seven parameters compared. While converting 2D future maps into 1D feature vectors, global max pooling produces the best performance in both the dataset while global average pooling has resulted in a poor performance. However, optimizer, temperature, alpha, and number of epochs are some of the least important hyperparameters. On the HAM10000 dataset, Adafactor has exhibited superior performance, outperforming RSMprop, which achieved the least accuracy. However, on the Kaggle dataset, RSMprop has emerged as the best optimizer among the four optimizers compared. Due to the conflicting outcome, Adafactor has been selected as the optimizer due to its faster convergence \cite{harrison2022closer}. Another conflicting outcome is produced with the temperature hyperparameter. On the HAM10000 dataset, the best classification performance is produced with temperature 1, and temperature 5 results in the worst performance. On the other hand, the best performance is achieved with a temperature value of 5 on the Kaggle dataset. In the proposed model, the temperature is set to 1, which implies that the soft targets from the teacher model are passed without any scaling. On both datasets, an alpha value of 0.5, a learning rate of 0.002, and a batch size of 64 have produced the best performance. Notably, the learning rate has been one of the most influential hyperparameters in this experiment. In both datasets, a learning rate of 0.1 has created an overfitted model while reducing the learning rate has gradually improved the accuracy. 

\begin{table}[h]
    \centering
    \caption{Ablation study on HAM10000 dataset.}
    \begin{tabular}{p{2.9cm} p{2.5cm} lllll}
    \hline
        \textbf{Hyper-parameter} & \textbf{Value} & \textbf{Accuracy} & \textbf{Precision} & \textbf{Recall } & \textbf{F1-score} & \textbf{MCC} \\ \hline
        Optimizer & Adam & 0.7414 & 0.7400 & 0.7414 & 0.7398 & 0.4689 \\ 
        ~ & Adafactor & 0.7484 & 0.7690 & 0.7484 & 0.7543 & 0.4632 \\ 
        ~ & Nadam & 0.7369 & 0.7365 & 0.7369 & 0.7348 & 0.4648 \\ 
        ~ & RMSprop & 0.7239 & 0.8460 & 0.7239 & 0.7542 & 0.4201 \\ \hline
        Batch Size & 16 & 0.7574 & 0.7566 & 0.7574 & 0.7569 & 0.7569 \\ 
        ~ & 32 & 0.8088 & 0.8089 & 0.8088 & 0.8088 & 0.5687 \\ 
        ~ & 64 & 0.8243 & 0.8255 & 0.8243 & 0.8248 & 0.6005 \\ \hline
        Learning & 0.1 & 0.6745 & 1.0000 & 0.6745 & 0.8056 & 0.0000 \\ 
        Rate & 0.01 & 0.7893 & 0.8101 & 0.7893 & 0.7963 & 0.4972 \\ 
        ~ & 0.001 & 0.7414 & 0.7852 & 0.7414 & 0.7317 & 0.5257 \\ 
        ~ & 0.002 & 0.8003 & 0.8540 & 0.8003 & 0.8158 & 0.5056 \\ \hline
        Temperature & 1 & 0.8313 & 0.8434 & 0.8313 & 0.8353 & 0.5979 \\ 
        ~ & 2 & 0.8178 & 0.8243 & 0.8178 & 0.8203 & 0.5710 \\ 
        ~ & 5 & 0.8088 & 0.8109 & 0.8088 & 0.8097 & 0.5562 \\ 
        ~ & 10 & 0.8103 & 0.8118 & 0.8103 & 0.8110 & 0.5514 \\ \hline
        Alpha & 0.3 & 0.8053 & 0.8075 & 0.8053 & 0.8063 & 0.5510 \\ 
        ~ & 0.5 & 0.8148 & 0.8221 & 0.8148 & 0.8176 & 0.5647 \\ \hline
        Feature & Flattening & 0.8442 & 0.8472 & 0.8442 & 0.8454 & 0.6336 \\ 
        Conversion & Global Average Pooling & 0.8038 & 0.8235 & 0.8038 & 0.8103 & 0.5286 \\ 
        ~ & Global Max Pooling & 0.9875 & 0.9875 & 0.9875 & 0.9875 & 0.9716 \\ \hline
        Epoch & 30 & 0.9775 & 0.9780 & 0.9775 & 0.9776 & 0.9487 \\ 
        ~ & 50 & 0.7474 & 0.7460 & 0.7474 & 0.7460 & 0.4804 \\ 
        ~ & \textbf{70} & \textbf{0.9875} & \textbf{0.9876} & \textbf{0.9875} & \textbf{0.9875} & \textbf{0.9719} \\ 
        ~ & 100 & 0.8223 & 0.8307 & 0.8223 & 0.8254 & 0.5772 \\ \hline
    \end{tabular}
    \label{tab:ham-ablation}
\end{table}

The number of epochs is the final hyperparameter that has been selected for the ablation study. The analysis indicates that 70 epochs provide ideal performance; after that, accuracy starts to decrease. In summary, the constructed student model is trained with the following hyperparameters: optimizer - Adafactor, batch size - 64, learning rate - 0.002, temperature - 1, alpha - 0.5, feature conversion - global max pooling, and epochs - 70. This configuration, as specified by the ablation study, defines the optimal combination for achieving high-performance results.

\begin{table}[h]
    \centering
    \caption{Ablation study on Kaggle dataset.}
    \begin{tabular}{p{2.9cm} p{2.5cm} lllll}
    \hline
        \textbf{Hyper-parameter} & \textbf{Value} & \textbf{Accuracy} & \textbf{Precision} & \textbf{Recall } & \textbf{F1-score} & \textbf{MCC} \\ \hline
        Optimizer & Adam & 0.8061 & 0.8133 & 0.8061 & 0.8073 & 0.6118 \\
        ~ & Adafactor & 0.8091 & 0.8363 & 0.8091 & 0.8094 & 0.6448 \\ 
        ~ & Nadam & 0.8136 & 0.8151 & 0.8136 & 0.8132 & 0.6282 \\ 
        ~ & RMSprop & 0.8424 & 0.8442 & 0.8424 & 0.8420 & 0.6858 \\  \hline
        Batch Size & 16 & 0.8364 & 0.8368 & 0.8364 & 0.8365 & 0.6691 \\ 
        ~ & 32 & 0.7894 & 0.7930 & 0.7894 & 0.7904 & 0.5711 \\ 
        ~ & 64 & 0.8212 & 0.8218 & 0.8212 & 0.8204 & 0.6393 \\  \hline
        Learning & 0.1 & 0.5682 & 1.0000 & 0.5682 & 0.7246 & 0.0000 \\ 
        Rate & 0.01 & 0.8030 & 0.8539 & 0.8030 & 0.8060 & 0.6493 \\ 
        ~ & 0.001 & 0.8076 & 0.8320 & 0.8076 & 0.8093 & 0.6340 \\ 
        ~ & 0.002 & 0.8303 & 0.8403 & 0.8303 & 0.8292 & 0.6708 \\ \hline
        Temperature & 1 & 0.8197 & 0.8235 & 0.8197 & 0.8193 & 0.6433 \\ 
        ~ & 2 & 0.8121 & 0.8255 & 0.8121 & 0.8145 & 0.6221 \\ 
        ~ & 5 & 0.8333 & 0.8596 & 0.8333 & 0.8341 & 0.6911 \\ 
        ~ & 10 & 0.7864 & 0.7908 & 0.7864 & 0.7864 & 0.5772 \\ \hline
        Alpha & 0.3 & 0.8076 & 0.8077 & 0.8076 & 0.8076 & 0.6141 \\ 
        ~ & 0.5 & 0.8424 & 0.8556 & 0.8424 & 0.8425 & 0.6980 \\ \hline
        Feature & Flattening & 0.8318 & 0.8501 & 0.8318 & 0.8311 & 0.6827 \\ 
        Conversion & Global Average Pooling & 0.8212 & 0.8214 & 0.8212 & 0.8208 & 0.6408 \\ 
        ~ & Global Max Pooling & 0.9848 & 0.9850 & 0.9848 & 0.9849 & 0.9697 \\ \hline
        Epoch & 30 & 0.9833 & 0.9835 & 0.9833 & 0.9833 & 0.9667 \\ 
        ~ & 50 & 0.9810 & 0.9811 & 0.9810 & 0.9811 & 0.9563 \\ 
        ~ & \textbf{70} & \textbf{0.9894} & \textbf{0.9894} & \textbf{0.9894} & \textbf{0.9894} & \textbf{0.9787} \\ 
        ~ & 100 & 0.9860 & 0.9861 & 0.9860 & 0.9860 & 0.9680 \\ \hline
    \end{tabular}
    \label{tab:kaggle-ablation}
\end{table}

\subsection{Model Quantization}
Model quantization is a method that represents parameters and activations with a lower bit precision than the conventional 64-bit floating-point format, thereby reducing the memory footprint and computing demands of deep neural networks \cite{rokh2023comprehensive}. This procedure is essential for deploying models on devices with limited memory and processing capability, like smartphones or edge devices. While quantization increases efficiency, over-quantization can reduce model accuracy due to the loss of fine-grained data \cite{al2023unified}.

To prevent potential accuracy erosion due to excessive quantization in our investigation, the final model is carefully quantized to 16-bit floating-point precision. This method ensures that performance does not decline by finding a balance between reducing the size of the model and keeping important information. Notably, before quantization, the model size was 2.03 MB. However, after quantization, it has drastically reduced to 469.77 KB, proving the usefulness of the quantization technique in reducing the model while preserving predictive accuracy.
% Moreover, the incorporation of the ReLU activation function, which facilitates the creation of a sparse matrix, significantly minimized the performance reduction caused by quantization.

\section{Results}
\label{sec:results}
This section presents the results achieved through the proposed method, alongside a comparison with existing studies. 

\subsection{Experimental Setup}
We have conducted this experiment on Kaggle. Python has been used as the programming language along with 6 libraries. The libraries are Tensorflow for deep learning model creation, Matplotlib for visualization, scikit-learn for evaluation metrics, Pandas for data storing in tabular format, OS for file location configuration, and TensorFlow Lite for model quantization. Moreover, we have leveraged GPU P100 to accelerate the training process.

\subsection{Evaluation Metrics}
To evaluate the model's performance, we have considered five evaluation metrics including accuracy, precision, recall, F1-score and Matthews Correlation Coefficient (MCC). While accuracy presents the overall correctness of the model, precision measures the accuracy of positive predictions. Recall, on the other hand, the ratio of true positive predictions to the actual positives. Accuracy measures the harmonic mean of precision and recall, presenting a more balanced result. MCC considers true positive, true negative, false positive, and false negative values to asses the models's performance. Since this method incorporates all coordinates of the confusion matrix, MCC is considered one of the crucial evaluation metrics, particularly for imbalanced datasets \cite{hicks2022evaluation}.

\begin{equation}
Accuracy =\frac{TP+TN}{TP+PP+TN+FN}
\end{equation}
\begin{equation}
Precision =\frac{TP}{TP+FP}
\end{equation} 
\begin{equation}
Recall=\frac{TP}{TP+FN}
\end{equation}
\begin{equation}
 F1-score=\frac{2 \times  Precision  \times  Recall }{ Precision+  Recall }
\end{equation}
\begin{equation}
MCC=\frac{TP \times TN-FP \times FN}{\sqrt{(TP+FP)(TP+FN)(TN+FP)(TN+FN)}}
\end{equation}

For additional assessment, we have presented confusion matrix and Receiver Operating Characteristic (ROC) curve. These metrics present a thorough assessment of the performance of the proposed solution. 

\subsection{Results of the Teacher Model}
Although the results of the teacher model do not reflect the deployed model, they provide critical guidance for the student model's development. Therefore, achieving a noteworthy performance with the teacher model is necessary. On the HAM10000 dataset, the teacher model achieves a high accuracy of 93.96\%, precision of 94.03\%, recall of 93.96\%, F1 score of 93.98\%, and MCC of 87.35\%. Similarly, on the Kaggle dataset, the model produces an excellent result, with an accuracy of 97.12\%, precision of 97.15\%, recall of 97.12\%, F1 score of 97.12\%, and MCC of 94.24 percent.

\subsection{Results of the Student Model}
\begin{figure}[h]
    \centering
    \begin{minipage}[b]{0.45\textwidth}
        \centering
        \includegraphics[width=\textwidth]{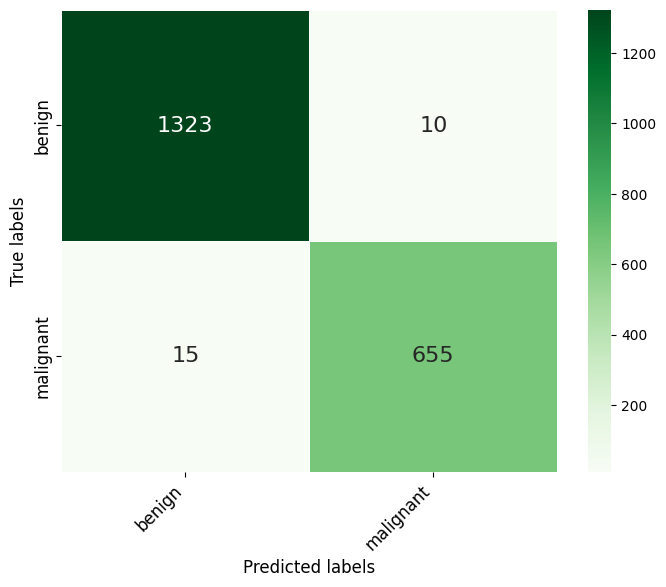}
        Confusion matrix of the student model on HAM10000 dataset.
    \end{minipage}
    \hfill
    \begin{minipage}[b]{0.45\textwidth}
        \centering
        \includegraphics[width=\textwidth]{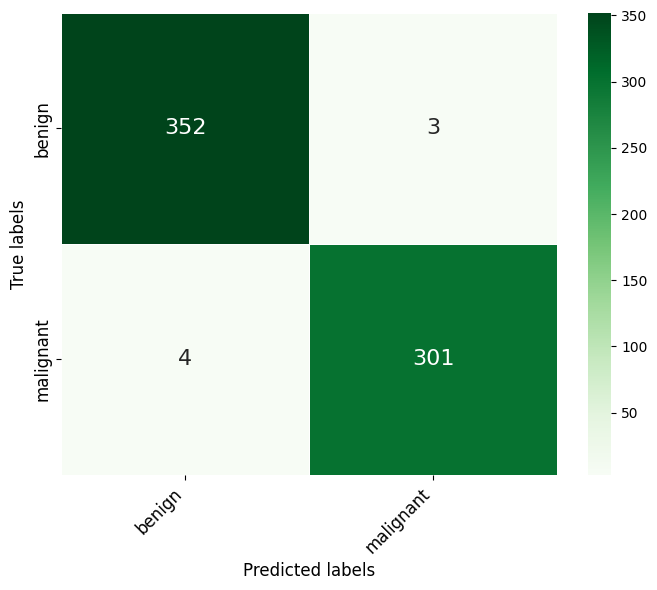}
        Confusion matrix of the student model on Kaggle dataset.
    \end{minipage}
    \caption{Confusion matrices of the student model.}
    \label{fig:cm}
\end{figure}

The student model's results demonstrate its remarkable performance in categorizing skin cancer images on both the HAM10000 and Kaggle datasets. On the HAM10000 dataset, the student model has achieved a remarkable accuracy of 98.75\%, precision of 98.76\%, recall of 98.75\%, F1 score of 98.75\%, and MCC of 97.19\%. Similarly, on the Kaggle dataset, the student model has achieved an exceptional accuracy of 98.94\%. Likewise, a precision, recall, and F1-score of 98.94\%, and MCC of 97.87\%. For further insight into the performance, the confusion metrics of the model are presented in Figure \ref{fig:cm}. 

\begin{figure}[h]
    \centering
    \begin{minipage}[b]{0.45\textwidth}
        \centering
        \includegraphics[width=\textwidth]{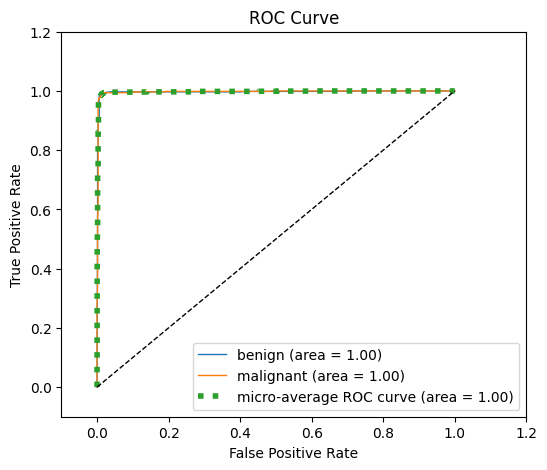}
        ROC curve of the student model on HAM10000 dataset.
    \end{minipage}
    \hfill
    \begin{minipage}[b]{0.45\textwidth}
        \centering
        \includegraphics[width=\textwidth]{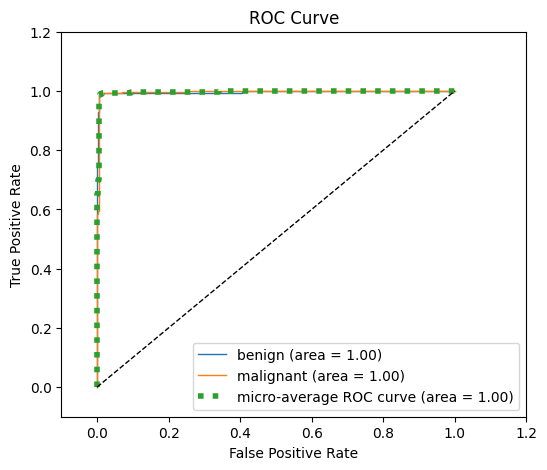}
        ROC curve of the student model on Kaggle dataset.
    \end{minipage}
    \caption{ROC curve of the student model.}
    \label{fig:roc}
\end{figure}

The figure depicts that among the 1333 benign images in the HAM1000 dataset, only 10 were misclassified as malignant, accounting for less than 1\% of the total benign images. A similar result is observed on the Kaggle dataset where only 1.1\% benign type cancers are misclassified. For the malignant type cancers, the misclassification rate is 2.2\% and 1.3\% on the HAM10000 and Kaggle datasets respectively. The ROC curves, presented in Figure \ref{fig:roc}, illustrate the model's exceptional performance in distinguishing between true positive and false positive classifications. Notably, the model scored an ROC value of 1.0 in all classes, demonstrating its exceptional performance and robustness across different skin cancer types. These remarkable results exhibit the usefulness of the proposed model in differentiating between benign and malignant skin lesions with high accuracy.  

\subsection{Comparison with Existing Works}
The majority of the research works have leveraged pretrained convolutional neural networks for classification. Among the well-known classifiers, EffecientNet and DenseNet are the most commonly used classifiers in classifying benign and malignant cancers \cite{venugopal2023deep, zia2022classification}. Through accurate fine-tuning and hyperparameter tuning, these models exhibit a noteworthy classification accuracy. Carcagni et al. \cite{carcagni2019classification} presented a classifier that integrates Support Vector Machine (SVM), a machine learning classifier, with DenseNet, a deep neural network. The modified network is more explainable and resource efficient. However, it failed to achieve a noticeable accuracy. 

\begin{table}[h]
    \centering
    \caption{Comparison of different methods on HAM10000 dataset}
    \begin{tabular}{ p{5cm} l l l l }
    \hline
        \textbf{Method} & \textbf{Accuracy} & \textbf{Precision} & \textbf{Recall} & \textbf{F1-score} \\ \hline
        EfficientNetV2-M \cite{venugopal2023deep} & 0.9595 & 0.83 & 0.94 & 0.88 \\ 
        Custom deep neural network \cite{ali2021enhanced} & 0.9016 & 0.9463 & 0.9391 & 0.9427 \\ 
        DenseNet with SVM \cite{carcagni2019classification} & 0.8525 & 0.91 & 0.9175 & 0.9137 \\ 
        Proposed & 0.9875 & 0.9876 & 0.9875 & 0.9875 \\ \hline
    \end{tabular}
    \label{tab:ham-compare}
\end{table}
Due to the large size of state-of-the-art image classifiers, some researchers have experimented with custom shallow neural networks which have produced impressive accuracy \cite{ali2021enhanced, montaha2022shallow, ghosh2022skinnet}. A comparative analysis of the proposed model with other research works on HAM10000 and Kaggle dataset is presented in Table \ref{tab:ham-compare} and \ref{tab:kaggle-compare} respectively. According to the analysis, the proposed solution achieves a superior performance over the majority of the research works. 

\begin{table}[h]
    \centering
    \caption{Comparison of different methods on Kaggle dataset}
    \begin{tabular}{ p{5cm} l l l l }
    \hline
        \textbf{Method} & \textbf{Accuracy} & \textbf{Precision} & \textbf{Recall} & \textbf{F1-score} \\ \hline
        Modified DenseNet201 \cite{zia2022classification} & 0.955 & 0.9702 & 0.9696 & 0.9546 \\ 
        Custom shallow CNN \cite{montaha2022shallow} & 0.9874 & 0.9856 & 0.9927 & 0.997 \\ 
        SkinNet-16 \cite{ghosh2022skinnet} & 0.9919 & - & - & - \\ 
        Proposed & 0.9894 & 0.9894 & 0.9894 & 0.9894 \\ \hline
    \end{tabular}
    \label{tab:kaggle-compare}
\end{table}

The superiority of the proposed system can be attributed to three major reasons. Firstly, the proposed solutions predominantly leverage CNNs which are only limited to local context. The proposed solution utilizes a teacher model composed of ViT and CNN to guide the student model. Thereby, the student model imitates the behavior of the teacher model which is aware of both the global and local context. Secondly, the majority of the works do not integrate any image preprocessing techniques. Since skin cancer images often contain hairs and other unwanted noise, a thorough image preprocessing technique significantly uplifts the image quality. However, the suggested approach integrates a six-stage image preprocessing technique, resulting in better classification performance. Lastly, unlike the presented method, some research works lack ablation study, leaving the impact of the hyper-parameters unknown which often leads to lower accuracy. The integration of the aforementioned approaches exalts the system's superiority over other works. Furthermore, very few research works focus on striking a balance between classification accuracy and model size, which makes the research work a unique contribution to skin cancer classification.

\subsection{Discussion}
This work employs a robust teacher model to guide the lightweight student model. Although the size of the teacher model is 975.85 MB, its sole purpose is to present soft instructions to the student model while training. The model has achieved an accuracy of 93.96\% and 97.12\% on HAM1000 and Kaggle datasets respectively. However, the lightweight student model has outperformed the teacher model by 4.79\% on the HAM1000 dataset and 1.82\% on the Kaggle dataset. The exceptional performance of the student model over the teacher model has two main causes. Firstly, while the teacher model only learns from the cross-entropy loss, the student model learns from both the cross-entropy loss and the distillation loss. Secondly, since the teacher model is not deployed for skin cancer classification, no ablation study is conducted, creating a gap in understanding the impact of hyperparameters. The student model, however, underwent a rigorous ablation study, significantly enhancing its classification capability.

\section{Conclusion}
\label{sec:conclusion}
In this article, we have presented a lightweight skin cancer classifier that achieves significant classification accuracy in differentiating benign and malignant skin cancer. The offered solution involves a six-step image preprocessing technique along with various data augmentation methods that enrich the training image. Subsequently, a robust teacher model is constructed to guide the lightweight student model for the final classification. With a comprehensive ablation study, the best hyperparameter configuration is achieved. The student model is further compressed with 16-bit quantization, reducing the size of the model to only 469.77 KB. We have assessed the presented solution on two benchmark datasets (HAM10000 and Kaggle). The proposed model has achieved an exceptional 98.75\% accuracy on the HAM10000 dataset and 98.94\% accuracy on the Kaggle dataset. A comparative study showcases the superiority of the proposed solution. This research, however, only considers binary classification. Future studies can explore the extension of this approach to multiclass classification tasks. Additionally, explainable AI techniques can be employed to analyze the model's predictability, providing insights into the decision-making process and increasing the classifier's transparency and trustworthiness. Overall, the suggested solution makes a substantial contribution to skin cancer diagnosis, especially in places with limited computational resources, by providing a lightweight yet accurate model for skin cancer classification. Its potential application in such situations emphasizes its importance and impact on improving healthcare accessibility.

\section*{Funding}
This research work received no external funding.

\section*{Data Availability Statement}
The HAM10000 dataset is available at \url{https://doi.org/10.7910/DVN/DBW86T} (accessed on 1st March 2023) and the Kaggle dataset is available at \url{https://www.kaggle.com/datasets/fanconic/skin-cancer-malignant-vs-benign} (accessed on 1st March 2023).

\section*{Conflicts of Interest}
The authors declare no conflict of interest.

\printbibliography
\end{document}